\documentclass{PoS}
\usepackage{amsmath}
\usepackage[utf8]{inputenc}

\makeatletter
\newcommand{\oset}[3][0ex]{%
  \mathrel{\mathop{#3}\limits^{
    \vbox to#1{\kern-2\ex@
    \hbox{$\scriptstyle#2$}\vss}}}}
\makeatother

\title{Implementing IceCube in SNOwGLoBES}

\ShortTitle{IceCube in SNOwGLoBES}

\author{
The IceCube Collaboration\footnote{For collaboration list, see PoS(ICRC2019) 1177.}\\
{\itshape \href{http://icecube.wisc.edu/collaboration/authors/icrc19_icecube}{http://icecube.wisc.edu/collaboration/authors/icrc19\_icecube}}\\
E-mail: \email{felixm@kth.se, erin.osullivan@fysik.su.se}
}

\abstract{We present an implementation of IceCube in the SNOwGLoBES package, which is used to calculate expected detection event rates resulting from supernova neutrinos. The SNOwGLoBES package is widely used to compare the sensitivity of different neutrino observatories, but currently does not include simulation files for IceCube. In this paper, we give a brief overview of the design process that went into this implementation.

\vspace{4mm}
{\bfseries Corresponding authors:}
Felix Malmenbeck$^{1}$, \speaker{Erin O'Sullivan}$^{2}$\\
{$^{1}$ \itshape Royal Institute of Technology (KTH)}\\
{$^{2}$ \itshape Oskar Klein Centre and Dept. of Physics, Stockholm University}

}

\FullConference{36th International Cosmic Ray Conference -ICRC2019-\\
		July 24th - August 1st, 2019\\
		Madison, WI, U.S.A.}

\makeatother

\begin{document}

\section{Introduction}

\subsection{SNOwGLoBES}

SNOwGLoBES (SuperNova Observatories with GLoBES) \cite{SGmanual}
is a software package that is used to calculate estimated neutrino
detection rates in the event of a supernova. As the name implies, SNOwGLoBES makes use of the GLoBES package, which is widely used for the simulation of a wide range of neutrino experiments on distance scales ranging from a few kilometers and up to that of solar neutrinos \cite{globes1, globes2}. SNOwGLoBES was developed 
with the aim of allowing fast, resource-efficient calculations, and has gained widespread use as a tool for comparing the response of different neutrino detectors to neutrinos from core collapse supernovae. With IceCube being the world's largest neutrino observatory, facilitating its inclusion in such comparisons is well-motivated.

Once a detector has been successfully added to the SNOwGLoBES package, users can supply their own neutrino flux spectra to calculate the response they would be expected to produce. Figure 1 shows one such calculation, using our implementation of IceCube in SNOwGLoBES to compare the neutrino signal for supernovae with varying progenitor masses.

\noindent 
\begin{figure}[h]
\noindent \begin{centering}
\includegraphics[width=1\columnwidth]{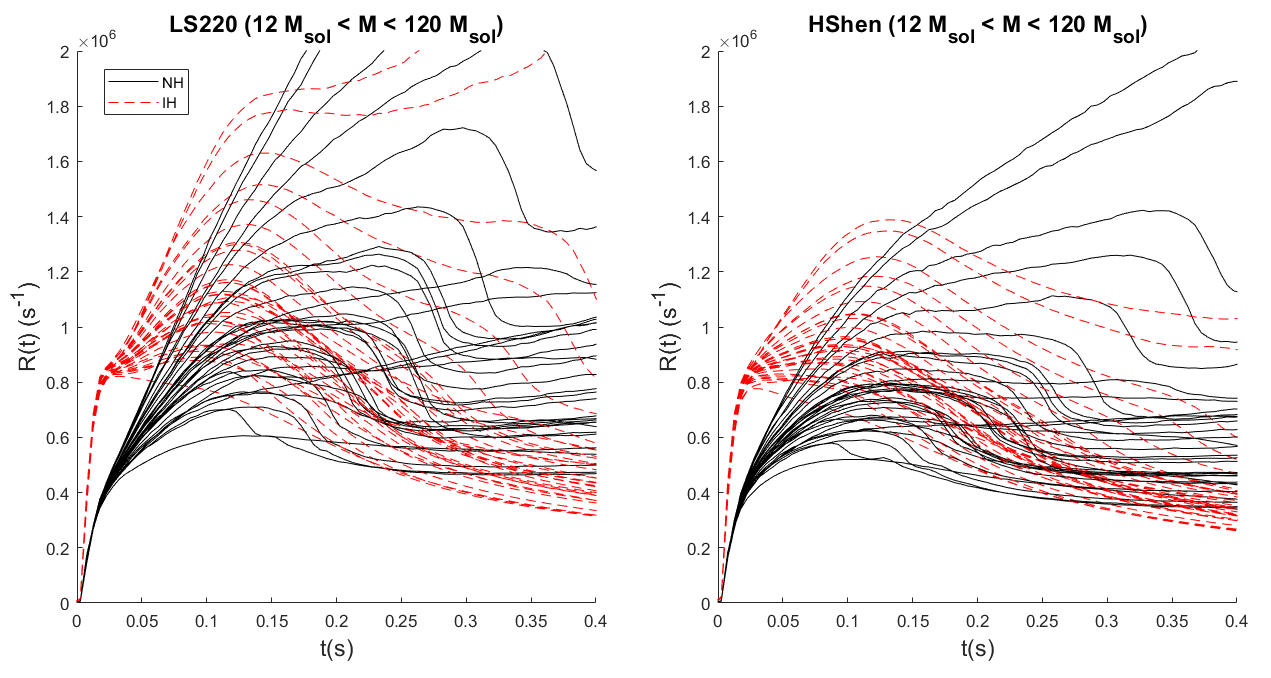}
\par\end{centering}
\caption{Comparing the time evolution of IceCube signals for 20 different supernova
progenitor masses, assuming either the Lattimer-Swesty (\textbf{left})
or HShen (\textbf{right}) equation of state. The solid (dashed) lines
show expected detection rates assuming the normal (inverted) hierarchy.
Ingoing fluxes were provided by \cite{lscomp}, and the expected detection rates were calculated using SNOwGLoBES.}
\end{figure}

\subsection{IceCube}

IceCube is a cubic-kilometer neutrino detector installed in the ice
at the geographic South Pole between depths of 1450 m and 2450 m,
completed in 2010. The ice is instrumented with 5160 digital optical
modules (DOMs).

IceCube's sensitivity to supernova neutrino signals has been the subject
of numerous studies, including \cite{lowenergy} and \cite{icrcSN}. As a (frozen) water Cherenkov detector, IceCube is primarily sensitive to the inverse beta decay (IBD) of $\bar{\nu}_e$ produced during the accretion phase of the supernova. The IBD interactions produce positrons in the detector volume, which in turn produce Cherenkov light detected by the DOMs. Other interaction channels, most notably neutrino-electron scattering, also contribute to this signal \cite{lowenergy,icrcSN,Kowarik}.
Although IceCube's sensitivity to low-energy neutrinos is not sufficient
to reconstruct individual supernova neutrino interaction
events, neutrino signals from supernovae occurring within our own galaxy are still expected
to show a significant increase in the DOM activity relative to random
noise, sufficient to study the time evolution
of the neutrino signal in detail.

\section{Theory}

\subsection{Calculating event rates in IceCube}

The rate of detections in IceCube due to a given interaction $I\,:\,\oset[0.2em]{(-)}{\nu}_{\mathrm{a}}+X\to e^{\mp}+Y+...$
can be approximated using the following formula, adapted from \cite{lowenergy}:

\begin{equation}
R_{\mathrm{a}}(t)=N_{\mathrm{DOM}}(t)\epsilon_{\mathrm{noise}}(t)\frac{n_{\mathrm{target}}L_{\mathrm{SN}}^{\mathrm{\nu,a}}(t)}{4\pi d^{2}\langle E_{\mathrm{\nu,a}}\rangle(t)}\int_{0}^{\infty}dE\int_{0}^{\infty}dE^{\prime}\frac{d\sigma_{\mathrm{I}}}{dE^{\prime}}(E^{\prime},E)N_{\mathrm{\gamma}}(E^{\prime})V_{\mathrm{\gamma}}^{\mathrm{eff}}f(E,t)
\end{equation}

Here,
\begin{itemize}
\item $R_{\mathrm{a}}(t)$is the number of detected interaction events per
second caused by supernova (anti)neutrinos of flavor $a$
\item $N_{\mathrm{DOM}}(t)$ is the number of DOMs collecting useable data
at time $t$
\item $\epsilon_{\mathrm{noise}}(t)\leq1$ is the deadtime efficiency factor (see Section 3.2)
\item $n_{\mathrm{target}}$ is the density of targets $X$ for a given reaction
\item $L_{\mathrm{SN}}^{\mathrm{\ensuremath{\nu},a}}(t)$ is the luminosity
of (anti)neutrinos of the flavor $a$
\item $d$ is the distance to the supernova
\item $\langle E_{\mathrm{\nu,a}}\rangle(t)$ is the average energy of incoming
(anti)neutrinos of flavor $a$
\item $d\sigma_\mathrm{I}/dE^{\prime}(E^{\prime},E)$ is the partial
cross-section for the given interaction resulting in an (anti)electron
with energy $E^{\prime}$
\item $N_{\mathrm{\gamma}}(E^{\prime})$ is the average number of photons
produced by an (anti)electron with energy $E^{\prime}$
\item $V_{\mathrm{\gamma}}^{\mathrm{eff}}$ is the effective volume of a single
photon in IceCube
\item $f(E,t)$ is the energy spectrum of incoming (anti)neutinos of flavor
$a$ with energy $E$ at time $t$
\end{itemize}

\subsection{IceCube effective volume}

When describing the sensitivity of a given neutrino detector to a core collapse supernova, it is common
to indicate its fiducial mass (or fiducial volume), as a means of describing how much
detector material is available for detection to take place. This is also the case in SNOwGLoBES, where each detector is assigned a given fiducial mass.

This fiducial mass is normally closely tied to the physical mass of the detector
material, making it relatively simple to visualize and estimate. However, in the case of IceCube, the array's great size and lack of
well-defined borders makes the fiducial mass and volume of the detector somewhat
ambiguous. IceCube has 5160 DOMs spread out over a volume of about
$1\,\mathrm{km}^{3}$. However, as is illustrated in Figure 2, the likelihood
that a given event will be observed drops off with its distance to
the nearest DOM, as well as the opacity of the ice in which it occurs.
The distance from which an interaction event can be detected is also
dependent on the energy of the incoming neutrino. As such, a high-energy
cosmic ray neutrino may produce a signal across multiple detectors,
allowing for the identification and reconstruction of single events,
whereas the light produced by interactions with supernova neutrinos often fades below
the detection threshold without triggering a single detector \cite{lowenergy}.
\noindent \begin{center}
\begin{figure}[h]
\begin{centering}
\includegraphics[height=60mm]{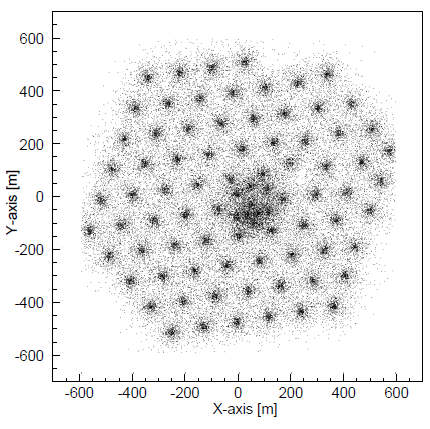}\includegraphics[height=62mm]{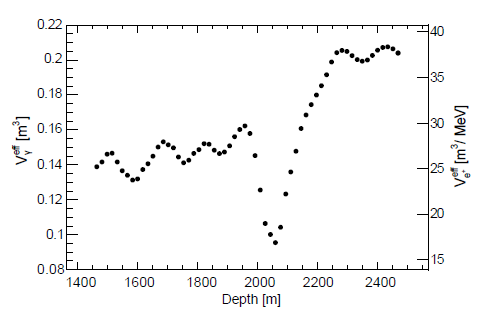}
\par\end{centering}
\caption{Figures from \cite{lowenergy}, illustrating how IceCube's sensitivity
to low-energy neutrino interactions varies with the location of the
event.
\textbf{Left: }The spatial distribution of detected supernova neutrino
interaction events, simulated using GEANT-3.21. Each dot represents
a interaction event which was detected by a digital optical module. \textbf{Right: }The effective volume $V_{\mathrm{\ensuremath{\gamma}}}^{\mathrm{eff}}$ for detecting Cherenkov photonswith wavelength (300 - 600) nm as a function of depth in the ice.}
\end{figure}
\par\end{center}

Simulating this behavior on an event-by-event basis can be quite complicated
and resource-intensive. To get around this issue, we use a concept
known as the effective volume. The effective volume simplifies calculations
of DOM hit rates by treating a large, imperfect detector of volume
$V$ through the analogy of a smaller detector of volume $V_{\mathrm{eff,tot}}$
wherein every interaction event results in a detection event.

When calculating the effective volume per DOM, we first calculate the rate at which supernova neutrinos give rise to the production of detectable photons within the detector volume, primarily in the form of Cherenkov radiation from (anti)electrons produced or scattered through inverse beta decay and neutrino-electron scattering. In making this calculation, we have followed the process outlined in \cite{Kowarik}, and arrived at the following expression for the total effective volume of IceCube with respect to (anti)electrons with energy $E^{\prime}$:

\begin{equation}
V_{\mathrm{eff,tot}}(E^{\prime},t)=N_{\mathrm{DOM}}(t)\times\epsilon_{\mathrm{noise}}(t)\times N_{\mathrm{\ensuremath{\gamma}}}(E^{\prime})\times V_{\mathrm{\ensuremath{\gamma}}}^{\mathrm{eff}}=N_{\mathrm{DOM}}(t)\times\epsilon_{\mathrm{noise}}(t)\times\langle V_{\mathrm{eff,\ensuremath{\pm}}}\rangle(E^{\prime}),
\end{equation}
\\
where $\langle V_{\mathrm{eff},\pm}\rangle(E^{\prime})$ is the average
effective volume of an (anti)electron produced with energy $E^{\prime}$:

\begin{equation}
\langle V_{\mathrm{eff,\ensuremath{\pm}}}\rangle(E^{\prime})=\theta(E^{\prime}-E_{\mathrm{ch}})\times(E^{\prime}-E_{\mathrm{ch}})\times C_{\mathrm{\ensuremath{\pm}}}\times\frac{dN_{\mathrm{\ensuremath{\gamma}}}}{dx}\times\langle V_{\mathrm{eff,\ensuremath{\gamma}}}\rangle,
\end{equation}
\\
where:
\begin{itemize}
\item $\theta$ is the Heaviside step function.
\item $E_{\mathrm{ch}}$ is the Cherenkov energy threshold, $E_{\mathrm{ch}}=m_{\mathrm{e}}+0.272\,[MeV]=0.783\,[MeV]$.
\cite{Kowarik}
\item The factor $C_{\mathrm{\ensuremath{\pm}}}$ quantifies a small difference
in the behavior of positrons and electrons, with values $C_{+}=0.577\left[\mathrm{\ensuremath{\frac{cm}{MeV}}}\right]$
and $C_{-}=0.580\left[\mathrm{\ensuremath{\frac{cm}{MeV}}}\right]$, as detailed in \cite{Kowarik}.
\item $dN_{\mathrm{\ensuremath{\gamma}}}/dx=325.35\,[\mathrm{c\ensuremath{m^{-1}}}]$
is the rate at which photons with wavelengths between $300\,[\mathrm{nm}]$
and $600\,[\mathrm{nm}]$ are generated as an (anti)electron moves
through the ice.
\item $\langle V_{\mathrm{eff},\gamma}\rangle=0.1575\,[\mathrm{m^{3}}]$ is the
mean effective volume for photons in the IceCube ice (c.f. Figure
3, right), which has been calculated through simulations, as detailed in \cite{Kowarik}.
\end{itemize}

\subsection{Event rates in SNOwGLoBES}

In SNOwGLoBES, calculated detection rates can be expressed using the
following formula (modified from the GLoBES user documentation \cite{globes1,globes2}):

\begin{equation}
R_{I}(t)=M_{\mathrm{detector}}n_{\mathrm{weight,I}}(\Delta E)^{2}\underset{j,k=1}{\overset{B}{\sum}}F_{\mathrm{a}}(E_{\mathrm{j}})\sigma_{\mathrm{I}}(E_{\mathrm{j}})k_{\mathrm{I}}(E_{\mathrm{k}},E_{\mathrm{j}})T_{\mathrm{I}}(E_{\mathrm{k}})
\end{equation}

Here,
\begin{itemize}
\item $R_{\mathrm{I}}(t)$ is the number of detections resulting from interaction
$I$
\item $M_{\mathrm{detector}}$ is the fiducial mass of the detector material
\item $n_{\mathrm{weight,I}}$ is the number of interaction targets per reference
target (see Section 3.1)
\item $\Delta E=E_{\mathrm{j}}-E_{\mathrm{j-1}}$ is the width of a single energy
bin
\item $B$ is the number of energy bins
\item $F_{\mathrm{a}}(E_{\mathrm{j}})$ is the flux of supernova (anti)neutrinos
of flavor $a$ with energy $E_{\mathrm{j}}$
\item $k_{\mathrm{I}}(E_{\mathrm{k}},E_{\mathrm{j}})$ is the energy distribution
function describing the proportion of interaction products produced
with energy $E_{\mathrm{k}}$
\item $T_{\mathrm{I}}(E_{\mathrm{k}})$ is the post-smearing efficiency function,
describing the probability that a particle with the energy $\mathrm{\ensuremath{E_{k}}}$
produced in interaction $I$ will result in a detection event
\end{itemize}

\section{Implementation}

\subsection{Channels and cross-sections}

SNOwGLoBES expresses the differential cross section as a product of
the total cross section $\sigma_{I}(E)$ and an energy resolution
function $k_{f}(E^{\prime},E)$:

\begin{equation}
\frac{d\sigma_{\mathrm{I}}}{dE^{\prime}}(E^{\prime},E)=\sigma_{\mathrm{I}}(E)k_{\mathrm{f}}(E^{\prime},E)
\end{equation}

The channels considered in this implementation of IceCube are the
same as those which are considered in the Water Cherenkov experiment
files which come pre-packaged with SNOwGLoBES v1.2 \cite{SGmanual} (see Table 1).
Likewise, we have made use of the pre-packaged total cross section
files for all channels. For the interactions involving oxygen atoms,
the pre-packaged files have also been used for the post-smearing efficiency functions. For the inverse
beta decay and electron scattering channels, the respective post-smearing efficiency functions have
been generated using the sources listed in Table 1, selected to agree
with the cross-sections referred to in \cite{lowenergy}.

Each channel has a weighting factor describing the relative number density of interaction
targets compared to a reference target. Here, the reference target is defined
to be hydrogen nuclei, which means that each water molecule (consisting of one oxygen nucleus, two hydrogen nuclei and 10 electrons)
contains two reference targets.

\begin{table}[h]
\noindent \begin{centering}
\begin{tabular}{|c|c|c|}
\hline 
Interaction & $n_{weight}$ & Partial cross sections\tabularnewline
\hline 
\hline 
$\bar{\nu}_{\mathrm{e}}+p\to e^{+}+n$ & 1 & Strumia, 2003 \cite{strumia}\tabularnewline
\hline 
$\nu_{\mathrm{e}}+e^{-}\to\nu_{e}+e^{-}$ & 5 & Marciano, 2003 \cite{marciano}\tabularnewline
\hline 
$\bar{\nu}_{\mathrm{e}}+e^{-}\to\bar{\nu}_{e}+e^{-}$ & 5 & Marciano, 2003 \cite{marciano}\tabularnewline
\hline 
$\nu_{\mu+\tau}+e^{-}\to\nu_{\mu+\tau}+e^{-}$ & 5 & Marciano, 2003 \cite{marciano}\tabularnewline
\hline 
$\bar{\nu}_{\mu+\tau}+e^{-}\to\bar{\nu}_{\mu+\tau}+e^{-}$ & 5 & Marciano, 2003 \cite{marciano}\tabularnewline
\hline 
$\nu_{e}+^{16}\mathrm{O}\to e^{-}+\mathrm{X}$ & 0.5 & (pre-packaged) \cite{SGmanual}\tabularnewline
\hline 
$\bar{\nu}_{e}+^{16}\mathrm{O}\to e^{+}+\mathrm{X}$ & 0.5 & (pre-packaged) \cite{SGmanual}\tabularnewline
\hline 
$\nu_{\mathrm{all}}+^{16}\mathrm{O}\to\nu_{\mathrm{all}}+\mathrm{X}$ & 0.5 & (pre-packaged) \cite{SGmanual}\tabularnewline
\hline 
\end{tabular}
\par\end{centering}
\caption{Interaction channels contributing to the supernova neutrino signal
in IceCube. }
\end{table}

\subsection{Correlated noise}

Apart from random Poisson noise, there is also correlated noise,
which is described in greater detail in \cite{lowenergy}. There are different ways of mitigating the effect
of this correlated noise, and which one is used will affect the final
detection rate.  An example is seen in \cite{lowenergy}, where an
artificial deadtime $\tau=250\,[\mathrm{\ensuremath{\mathrm{\mu s}}}]$
is introduced after each DOM hit, result in the detection rate being
modified by a factor $\epsilon_{\mathrm{noise}}\approx0.87/(1+\tau\cdot r_{\mathrm{SN}}(t))$. However, it should be noted that recent advances such as the development of HitSpooling is expected to allow for more efficient handling of correlated noise, resulting in less loss of signal \cite{hitspool}.

At present, SNOwGLoBES does not have native support for time-dependent
or dynamic effects. In this work, therefore, a constant deadtime efficiency factor $\epsilon_{\mathrm{noise}}(t)\doteq0.95$ is used,
instead. Setting the built-in noise screening factor to be constant
has the benefit that users who implement their own noise screening
factors can easily divide by this constant to obtain the unscreened
detection rates.

The noise rates in the IceCube DOMs average 540 Hz \cite{lowenergy,icrcSN}, and the artificial deadtime reduced this rate by roughly 50\%. SNOwGLoBES does allow the user to include a pre-defined, constant
background noise profile. We have not made use of this function, as
adding noise using other tools provides greater flexibility and a
better simulation of both Poisson noise and correlated fluctuations.
However, adding a noise profile in the future may serve a pedagogical
function for users who are not familiar with IceCube's significant
rate of background noise.

\subsection{Detector mass and effective volume}

One important aspect of defining an experiment in SNOwGLoBES is determining
the experiment's detector mass, which acts as a constant scale factor.
At present, SNOwGLoBES does not natively support allowing this mass
to vary with the interaction channel or the energy of the interaction
products. As a result, we cannot define the experiment's detector
mass to correspond to the effective volume calculated in Section 2.2.

To get around this limitation, we have instead included the calculation
of effective volume in the post-smearing efficiency function, $T_{\mathrm{I}}(E_{\mathrm{k}})$.
In this work, we have set IceCube's fiducial mass to $M=51600\,\mathrm{kton}$,
corresponding to a volume of about $10^{4}\,\mathrm{\ensuremath{m^{3}}}$
per DOM. We then define the post-smearing efficiency function for
the interaction $I$ such that

\begin{equation}
T_{I}(E_{k})=\frac{\rho_{\mathrm{ice}}V_{\mathrm{eff,tot}}(E_{\mathrm{k}})}{M}=\frac{\rho_{\mathrm{ice}}N_{\mathrm{DOM}}(t)\times\epsilon_{\mathrm{noise}}(t)\times\langle V_{\mathrm{eff,\ensuremath{\pm}}}\rangle(E^{\prime})}{M}.
\end{equation}

For this calculation, we have assumed that $N_{DOM}(t)=0.98\times5160$,
meaning that 98\% of DOMs are collecting data that can be used for
analysis. This value was chosen the sake of consistency with \cite{lowenergy}.
An ice density of $\rho_{\mathrm{ice}}=910\,\mathrm{kg/m^{3}}$
has been used.

\section{Discussion}

Using the procedures described above, we have developed a working
implementation of IceCube in the SNOwGLoBES package, and submitted
it for incorporation into the standard distribution of the package\cite{githubpull}. The SNOwGLoBES software has developed into a standard tool for comparing the response of many different neutrino detectors to neutrinos from a core collapse supernova. Since IceCube is the world's largest neutrino observatory, incorporating its detector response into the software is well-motivated. The procedures that went into this development can be applied to other large water Cherenkov detectors, such as KM3Net, provided an estimate of the detector effective volume.

Instructions for installing and using SNOwGLoBES
can be found in the user manual \cite{SGmanual}.

\end{document}